# Achromatic diffractive lens limits

JACOB ENGELBERG AND URIEL LEVY*

*Department of Applied Physics, The Center for Nanoscience and Nanotechnology, The Hebrew University, Jerusalem 91904, Israel*
*ulevy@mail.huji.ac.il*

**Abstract:** Over the recent years, there have been many reports of achromatic metalenses and diffractive lenses. However, very few (if any) practical applications of such achromatic flat lenses have been demonstrated, which raises questions about the potential of these lenses to provide solutions for real world cases which involve broadband illumination. A recent paper placed limits on the performance of achromatic metalenses. However, is this limit also valid for a diffractive lens? Not necessarily so. In this paper we derive the limits on the performance of achromatic diffractive lenses. In particular, we show that achromatic diffractive lenses can cover a wide spectral range, limited only by loss of efficiency caused by manufacturing limitations related to feature depth and size. On the other hand, we show that achromatic diffractive lenses can provide near diffraction limited performance only at very low Fresnel numbers, i.e. they cannot provide large focusing power and broadband response simultaneously. We then go on to compare the limits of achromatic metalenses and diffractive lenses, in attempt to understand the potential of different types of flat lenses. Our findings may set the ground for better evaluation of flat lens performance, understanding of their capabilities and limitations, and for exploring novel design concepts and applications.

## 1. Introduction

Over the years, there has been a lot of interest in flat lenses, a category including two competing technologies: diffractive lenses [1–4] and the "new comer" - metalenses [5–13]. These lenses provide a compact and cost-effective solution that may replace conventional multi-element refractive lens designs for some applications, such as microscope objectives and cellphone camera lenses, and are thus of great interest to the community at large. A major obstacle in the path towards realizing the potential of such lenses is the large chromatic aberration associated with diffractive lenses (metalenses are actually a sub-category of diffractive lenses, so they exhibit the same basic dispersion). To meet this challenge, achromatic metalenses (AMLs) [14–21] and achromatic diffractive lenses (ADLs) [22–25] have been developed. While these designs have shown achromatic behavior, they have not so far shown much utility for practical applications. This raises a question regarding their potential to provide the performance needed to meet the requirements of modern optical systems.

In a recent paper, the upper limit of achromatic metalens performance was explored [26]. It was shown that there is an inverse relation between the bandwidth over which the AML operates and the time delay it can provide (the time delay represents the "work" the lens is doing, which we will show is related to the Fresnel number of the lens). In contrast to this, inverse-designed ADLs seem to suffer no such limitation. Recent papers have reported ADLs that operate over an extremely large wavelength bandwidth [27,28]. If so, what are the limitations of ADLs? This question has been raised by several authors [26–28], but to our knowledge has not been properly answered yet. In this paper we provide an answer to this question. Specifically, we show that ADLs suffer from a severe limitation, preventing them from providing near-diffraction limited performance at high numerical aperture (NA) and broadband illumination. On the other hand, at low NA they can indeed cover a very large spectral band. We also discuss the limits of the spectral band that can be covered.

Based on our analysis we compare the competing technologies of AMLs and ADLs. We show that ADLs have an advantage for low NA applications with wide spectral range. AMLs on the other hand have an advantage for high NA applications with modest spectral band. Both technologies still cannot provide near diffraction limited performance at high-NA and wide spectral band, which is required, for example, for such applications as cellphone cameras and microscope objectives.

If what we say is true, how do we explain the many reports of moderate or even high-NA ADLs with near diffraction limited performance [27–32]? The answer is that the real limitation is not on NA but rather on the Fresnel number (FN), as will be explained in the paper. For most practical applications, the lens dimension (aperture/focal length) should be orders of magnitude larger than the wavelength (e.g. 1mm focal length and 1µm wavelength). In these cases, the limitation of low FN translates to low NA. The reports mentioned above either have very small dimensions, or provide resolution which is much lower than the diffraction limit, but mask this fact by presenting the resolution in terms of the less relevant full width half maximum (FWHM) instead of more relevant criteria for evaluating imaging lenses such as MTF or Strehl ratio.

Based on our analysis one can gain significant insight into the physical mechanisms driving the different flat lens solutions, providing tools that can help find the best solution for a given application, and hopefully generate new knowledge based on novel solutions.

## 2. Achromatic metalens limit

In [19] an upper limit is set for the bandwidth of a dispersion engineered AML, that depends on the maximum dispersion range that can be obtained with a specific nanoantenna library. In [26] this limit is generalized by stating that the maximum dispersion range obtainable with any library will be between zero (i.e. effective refractive index independent of frequency) and $\Delta n = n_a - n_b$ (i.e. the effective refractive index is equal to that of the bulk nanoantenna material, $n_a$ for the highest frequency, and to the background index $n_b$, for the lowest frequency). This results in Tucker's upper limit for the time-delay-bandwidth product [33]:

$$\Delta T \Delta \omega \leq \Delta \varphi_{max} \qquad (1)$$

Where $\Delta \varphi_{max} = \frac{2\pi}{\lambda_0} h \Delta n$ is the maximum phase difference between the highest and the lowest frequency, $h$ is the nanostructure height, and $\lambda_0$ is the central wavelength (CWL) of the spectral range $\Delta \omega$. Since the necessary time delay $\Delta T$ for a lens of given parameters is known, the maximum bandwidth $\Delta \omega$ can be found. The delay-bandwidth terminology is natural for optical communications applications, but for imaging optics it is more convenient to express the maximum time delay needed in terms of the Fresnel number (FN) of the lens [26]:

$$\Delta T_{max} = \frac{OPD_{max}}{c} \approx \frac{FN \cdot \lambda_0/2}{c}, \quad OPD_{max} = \sqrt{f^2 - R^2} - f \qquad (2)$$

Where $OPD_{max}$ is the maximum optical path difference that a lens of focal length $f$ provides at its maximum aperture radius $R$, and $c$ is the speed of light in vacuum. The FN is defined according to Eq. 3, where $NA$ is the numerical aperture of the lens, and the paraxial approximation is used [34]:

$$FN = \frac{R^2}{\lambda_0 f} \approx \frac{f \cdot NA^2}{\lambda_0} \approx \frac{OPD_{max}}{\lambda_0/2} \qquad (3)$$

We can express the maximum bandwidth of the lens in terms of the FN, by substituting Eq. 2 into Eq. 1. Following this we can reverse the equation to express the maximum achievable FN for a given bandwidth. This results in Eq. 4, where $p$ is the height of the nanostructure in units of the central wavelength $\lambda_0$ ($p = h/\lambda_0$). The larger the relative spectral range, the lower the maximum FN that can be achieved. In addition, the larger the nanostructure height and its refractive index contrast with respect to its surrounding, the larger the FN that can be achieved.

$$FN_{max} = 2p\Delta n \left(\frac{\Delta\lambda}{\lambda_0}\right)^{-1} \tag{4}$$

## 3. Achromatic diffractive lens limit

Does the above delay-bandwidth limit apply to ADLs? In our opinion it does not. This is despite the fact that the propagation in ADLs is longitudinal (within the paraxial approximation) and not lateral, so the conditions mentioned in [26] seem to be unviolated. In the case of a dispersion engineered AML, the achromatic behavior is based on the dispersive nature of the truncated waveguides. Therefore, the analogy to the world of optical communications and delay lines is relevant. However, in the case of an ADL, the achromatic behavior is based on the variation in energy distribution among the various diffraction orders for different wavelengths, caused by the overall microscopic surface structure, rather than by a local effect related to a single point on the aperture. Therefore, the delay line analogy does not apply.

To understand what happens in the case of an ADL, we should consider the case of a multi-order diffractive (MOD) lens [22,23], which is the forerunner of modern inverse-designed ADLs [35]. A MOD lens is a diffractive lens that is designed to operate at a high diffraction order *m* for the nominal design wavelength. In such a case, the harmonic wavelengths, given by equation 5, will focus with high efficiency at the same point as the design wavelength.

A MOD lens cannot be perfectly achromatic over a spectral range, like an AML, but rather only at discrete wavelengths. A MOD, like an AML, creates the same time delay for the different harmonic frequencies. This is achieved by applying a phase function that varies linearly with frequency, as does the AML [19]. However, as mentioned before, the change in the phase as a function of frequency is not provided by structural dispersion of the truncated waveguides, but rather by structural dispersion of the diffractive surface. This dispersion causes different harmonic frequencies to focus with optimal efficiency at different diffraction orders, according to Eq. 5 [22]. $\lambda_0$ is the design wavelength, for which the MOD operates at the design order *m*. $\lambda$ and *k* represent a different harmonic wavelength and the diffraction order in which a peak in diffraction efficiency is achieved for this wavelength, respectively. ω is the angular frequency associated with $\lambda$.

$$\lambda = \frac{m\lambda_0}{k} \Rightarrow k = \frac{m\lambda_0}{\lambda} = \frac{\omega m \lambda_0}{2\pi c} \tag{5}$$

The phase function varies linearly with the order of diffraction *k*, and will therefore vary with frequency (per Eq. 5), thus achieving the same time delay at discrete resonant frequencies. This is described by Eq. 6, where $\phi_0$ is the nominal phase function (for design wavelength $\lambda_0$ at design order *m*), and $\phi$ is the phase function "seen" by a wavelength $\lambda$ operating at order *k*.

$$\phi = \frac{k}{m}\phi_0 = \frac{\omega \lambda_0}{2\pi c}\phi_0 \tag{6}$$

What happens in between harmonic wavelengths? The energy of these non-harmonic wavelengths will be mostly divided between the two closest diffraction orders, so there is minimal loss of efficiency. However, the focus of these wavelengths will be shifted compared to that of the harmonic wavelengths. The maximum focal shift is given by Eq. 7 [22], where $\Delta\lambda_{har}$ is the wavelength difference between neighboring harmonic wavelengths (e.g. $\lambda_0$ and $\lambda$ of order $k = m + 1$).

$$\frac{\Delta f}{f} = \frac{\Delta\lambda_{har}}{\lambda_0} = \frac{1}{m+1} \tag{7}$$

What limits the performance of a MOD lens is this chromatic focal shift. For the lens to be diffraction limited the maximum focal shift must be smaller than the diffraction limited depth of focus. For the maximum focal shift, we will use half of $\Delta f$ given by Eq. 7, since half-way to the nearest resonant wavelength, more than 50% of the power will already shift to the nearby resonance. So, while there will be some power that is defocused by the full $\Delta f$, most of the power is only defocused by a maximum of $\Delta f/2$. The diffraction limit condition is therefore given by Eq. 8, where the right-hand side is the diffraction limited depth-of-focus [36]. This criterion is not intended to be exact, but rather to give an estimate of the maximum FN up to which we can expect to obtain near diffraction limited performance.

$$\frac{\Delta f}{2} = \frac{f}{2(m+1)} < \frac{\lambda_0}{2NA^2} \tag{8}$$

Substituting Eq. 3 into Eq. 8, we obtain the upper limit on the Fresnel number of a diffraction limited MOD lens:

$$FN < m + 1 \tag{9}$$

The design order $m$ is related to the zone depth $h$ (this is the feature height, analogous to the truncated waveguide height in an AML) by $m = h\Delta n/\lambda_0 = p\Delta n$. Substituting this into Eq. 9 we obtain the following condition, which is analogous to the condition of Eq. 4 for metalenses:

$$FN_{max} = p\Delta n + 1 \tag{10}$$

The remarkable thing about this condition is that it is independent of spectral range. The only limit on the achievable FN is the zone depth and refractive index contrast. This is because for a high order MOD lens (e.g. $m$~10), the chromatic blur is limited by the distance between neighboring harmonic wavelengths $\Delta\lambda_{har}$, rather than by the overall spectral range. Note that this limit is only relevant to a MOD lens, and not to a CDL operating at a single order of diffraction. In the case of a CDL we can operate over a spectral range smaller than $\Delta\lambda_{har}$, thus achieving near diffraction limited performance at FN higher than the limit given by Eq. 10, but this will be a quasi-monochromatic mode of operation [37]. The case of a CDL is analyzed in section 4.

The limit given by Eq. 10 is based on the chromatic aberration near the nominal wavelength $\lambda_0$ at the diffraction order $m$. For larger wavelengths, operating at lower diffraction order ($k \ll m$), the aberration will be larger (in Eq. 7, $m$ is replaced by $k$). However, the diffraction limited spot size will also be larger, which more than compensates for the increased aberration. For shorter wavelengths ($k \gg m$), the situation is reversed. The limit for wavelengths far from the nominal (both longer and shorter) comes out to be $FN < m + \lambda/\lambda_0$ (where the FN is still given at the nominal wavelength $\lambda_0$). For large wavelengths satisfying $\lambda/\lambda_0 > 1$, the maximal FN is now higher than that obtained from Eq. 10, so this is not a limiting factor. For shorter wavelengths satisfying $\lambda/\lambda_0 < 1$, the limit can go as low as $FN < m$. Assuming sufficiently large $m$, this is a negligible change which can be ignored.

If the maximum FN is not affected by the spectral range, what limits the spectral range a MOD can operate over? Let us look first at the upper limit for the wavelength. The maximum wavelength for which we can obtain 100% efficiency is that in which the MOD operates at the first order of diffraction, $\lambda_1 = m\lambda_0$. At longer wavelengths the efficiency drops according to [38]:

$$\eta = sinc^2(\lambda_1/\lambda - 1) \quad , \quad sinc(x) = sin(\pi x)/\pi x \tag{11}$$

If we allow up to 50% drop in efficiency as the criterion for maximum wavelength, we obtain:

$$\lambda_{max} \approx 1.8m\lambda_0 = 1.8p\Delta n\lambda_0 \tag{12}$$

Naively, one would expect that the maximal wavelength could be enormous. For example, for a MOD with $m$=100, and $\lambda_0 = 1.5\ \mu m$, the maximal wavelength is in the terahertz range. However, this is impractical for at least a couple of reasons. First, it is challenging to find materials with high transparency for such a broad spectral band. Furthermore, for such a high value of $m$, the scalar expression for diffraction efficiency (Eq. 11) breaks down, and lower efficiencies are obtained [39]. In addition, a high $m$ leads to a semi-refractive lens, where material dispersion effects become dominant. Lastly, a very high $m$ lens may no longer merit the description of being "flat". In practice, moderate values of $m$ should be chosen for reasons of efficiency and manufacturability, and thus the maximal wavelength is expected to be limited to the infrared.

Now we move to the lower wavelength limit. Here we are limited by transverse sampling resolution. To obtain reasonable efficiency we must have at least two phase-levels for each $2\pi$ phase induced in the shortest wavelength (this will result in 40.5% efficiency going to the desired order based on the scalar approximation [39]). If we can manufacture $N$ phase levels, this means [22]:

$$\lambda_{min} = 2h\Delta n/N \qquad (13)$$

If $\Delta$ is the minimum transverse feature which can be manufactured by our machine, and assuming equally spaced phase levels, we obtain $N = a/\Delta$, where $a$ is the minimum period. For a MOD lens with a certain NA (at the nominal wavelength) $a$ is given by the grating equation [38]: $a = m\lambda_0/NA$. Substituting the last two expressions into Eq. 13 results in:

$$\lambda_{min} = 2\Delta \cdot \text{NA} \qquad (14)$$

Incidentally, this is the same minimum wavelength that would be obtained for first order operation at a given NA and transverse resolution of $\Delta$, according to the Nyquist criterion [3,40].

The above limits were determined based on a MOD lens. Do they apply to an inverse designed ADL? An inverse designed ADL is designed numerically rather than analytically. The lens aperture is divided into rings, whose heights are optimization variables. These heights are constrained to discrete values in the range of zero to the designated maximum depth $h$. The merit function reflects the efficiency and resolution in some manner, which are then hopefully maximized by the optimization algorithm [29].

While optimization techniques are useful for finding an optimal working point based on a specific figure of merit, they cannot create new physics or work miracles. Therefore in the same manner that the AML limit derived in [26], based on an ideal hyperbolic phase function, was empirically shown to apply to an inverse designed AML, the ADL limit derived here can also be expected to apply to inverse designed ADLs. To validate this claim, we performed an empirical study based on published results, similarly to the methodology of [26]. The results are presented in section 0.

### 4. Chromatic flat lens limit

Having established upper limits for AMLs and ADLs, we can now assess their potential for improvement over a CDL. The chromatic focal shift of a CDL is given by [38]:

$$\frac{\Delta f}{f} = \frac{\Delta \lambda}{\lambda_0} \qquad (15)$$

Unlike the case of an ADL, here $\Delta\lambda$ is the full desired bandwidth, rather than the difference between two neighboring harmonic wavelengths. The actual longitudinal chromatic aberration will be half of this shift since the CDL is optimized for the center wavelength $\lambda_0$. To obtain diffraction-limited performance we require that the longitudinal aberration be smaller than the depth of focus, as in Eq. 8. This results in:

$$FN_{max} = \left(\frac{\Delta\lambda}{\lambda_0}\right)^{-1} \qquad (16)$$

Comparing this to Eq. 4, we can see that they are similar, but the AML has an additional factor, related to the dispersion of the truncated waveguide. Therefore, the AML FN upper limit can be greater than that of the CDL, if nanoantenna height and index contrast are sufficiently high. Comparing this limit to that of an ADL, we see a major difference in that the ADL limit is independent of spectral range and is limited only by the profile height and index contrast. This ADL behavior is only relevant for spectral ranges broader than that of two neighboring harmonic wavelengths. For narrower spectral ranges it will function as a CDL, which can allow near diffraction limited performance at high Fresnel numbers, for very narrow spectral ranges.

## 5. Validation

To validate our ADL upper limits, we performed a survey of published ADL designs. The survey is not meant to be exhaustive and it does not include all ADLs demonstrated so far. However, it is more than sufficient to provide an understanding of the underlying physics and design considerations. The results of our survey are presented in Table 1. The table compares the following: FN (actual lens Fresnel no.) to $FN_{max}$ (maximum diffraction limited Fresnel no., based on Eq. 10), $\lambda_{min}$ (actual design minimum wavelength) to '$\lambda_{min}$ lim.' (minimum wavelength calculated according to Eq. 14) and $\lambda_{max}$ (actual design maximum wavelength) to '$\lambda_{max}$ lim.' (maximum wavelength calculated according to Eq. 12). These column headings are marked with light green background and bold letters.

The first two designs are MOD lenses. They were manufactured by diamond turning, so the effective feature size and number of phase levels are unknown. These parameters affect only the short wavelength limit, so we simply chose reasonable values (marked in orange). The rest of the designs are inverse designed ADLs. The shown efficiency results are the reported simulated results, except for design no. 2, where the efficiency is a measured value, and is mostly due to silicon absorption in the operating spectrum.

Table 1. Published design parameters compared to parameter limits presented in this paper

| No. | Ref. | NA | $f$ [mm] | $\Delta n$ | $h$ [µm] | $\Delta$ [µm] | $N$ | **FN** | **$FN_{max}$** | **$\lambda_{min}$ [µm]** | **$\lambda_{max}$ [µm]** | **$\lambda_{min}$ lim. [µm]** | **$\lambda_{max}$ lim. [µm]** | Eff. |
|---|---|---|---|---|---|---|---|---|---|---|---|---|---|---|
| 1 | [22] | 0.1 | 28.8 | 0.494 | 23.4 | 1 | 100 | 552 | 21 | 0.5 | 0.64 | 0.21 | 20.8 | 0.90 |
| 2 | [24] | 0.32 | 1 | 2.4 | 50 | 1 | 100 | 10 | 12.4 | 7 | 14 | 0.63 | 216 | 0.75 |
| 3 | [25] | 0.05 | 1 | 0.65 | 2.4 | 3 | 100 | 4.2 | 3.6 | 0.45 | 0.75 | 0.30 | 2.81 | 0.88 |
| 4 | [25] | 0.18 | 1 | 0.65 | 2.6 | 1.2 | 100 | 55 | 3.8 | 0.45 | 0.75 | 0.43 | 3.04 | 0.47 |
| 5 | [41] | 0.20 | 0.063 | 0.65 | 2 | 1 | 100 | 4.6 | 3.3 | 0.47 | 0.67 | 0.4 | 2.34 | 0.81 |
| 6 | [41] | 0.36 | 0.155 | 0.65 | 10 | 4 | 100 | 5.6 | 2.6 | 3 | 5 | 2.88 | 11.7 | 0.86 |
| 7 | [41] | 0.81 | 0.002 | 0.65 | 1.6 | 0.35 | 100 | 4.1 | 2.5 | 0.56 | 0.8 | 0.57 | 1.87 | 0.70 |
| 8 | [31] | 0.37 | 19 | 0.65 | 10 | 8 | 100 | 292 | 1.7 | 8 | 12 | 5.94 | 11.7 | 0.43 |
| 9 | [31] | 0.45 | 8 | 0.65 | 10 | 8 | 100 | 192 | 1.7 | 8 | 12 | 7.2 | 11.7 | 0.65 |
| 10 | [31] | 0.37 | 19 | 2.4 | 10 | 8 | 16 | 292 | 3.4 | 8 | 12 | 5.94 | 43.2 | 0.71 |
| 11 | [27] | 0.075 | 1 | 0.65 | 2.6 | 3 | 100 | 8.7 | 3.6 | 0.45 | 0.85 | 0.45 | 3.04 | 0.89 |
| 12 | [27] | 0.075 | 5 | 0.65 | 10 | 6 | 100 | 3.6 | 1.8 | 0.5 | 15 | 0.9 | 11.7 | 0.76 |
| 13 | [27] | 0.1 | 10 | 2.4 | 12 | 6 | 100 | 1.3 | 1.4 | 2.5 | 150 | 1.2 | 51.8 | 0.91 |
| 14 | [30] | 0.17 | 25 | 0.65 | 2.6 | 3 | 100 | 579 | 2.3 | 0.875 | 1.675 | 1.02 | 3.04 | 0.91 |

Let us first look at the operating wavelength ranges. Almost all the designs are within the theoretical limits, or very nearly so. Only designs 12 and 13, which attempt to push the wavelength limits to obtain a very broad spectral range, significantly exceed the limits (cells marked in pink in the relevant columns). However, it is important to remember that these limits are not hard limits. They express the fact that beyond these wavelengths the efficiency is expected to be low. For the short wavelength limit of design 12, a drop in efficiency is indeed reported (not shown in the table). For the long wavelength limits, the reported simulated efficiency is still high. If we take the example of design 13, the maximum "zone depth" in terms of phase for the longest wavelength (150nm) is $2\pi/\lambda \cdot h \cdot \Delta n = 1.2\ rad$. Even an ideal diffractive lens designed for first order operation at this wavelength but limited to this zone depth, would give an efficiency of less than 5%, as a result of the large detuning factor ($\alpha=1.2/(2\pi)=0.19$). How can this be reconciled with the very high (>90%) simulated efficiency [27]? This remains to be seen.

We now look at the Fresnel numbers of the reported designs. Most of the designs exceed the theoretical limit, some by a small amount, but others by a large amount – these are marked in pink. The theoretical limit reflects the maximum FN at which the lens can be diffraction limited, similar to the delay-bandwidth limit given in [26] for AMLs. Therefore, it should not come as a surprise that one can work at higher FNs. What we expect to see, however, is that at these higher FNs, the resolution will degrade accordingly. Unfortunately, most published resolution data relates only to FWHM, which is not an effective measure of resolution, as it ignores sidelobes or 'tails' of the point-spread function which can cause a severe drop in the image contrast/resolution, as represented by the modulation-transfer-function (MTF) [42]. Fortunately, there are a few publications that do provide MTF data. To make this data meaningful for our purposes, we compared the simulated/measured MTF data (the data was manually adapted from published data, so it is not very accurate, but sufficient to demonstrate the principle) to the diffraction limited MTF, which we simulated based on the provided lens parameters. From the examples shown in Fig. 1, corresponding to designs 1, 2, 8, and 11 of Table 1, it is clear that while it is possible to design ADLs whose FN exceeds the upper limit, the farther it is from upper limit, the lower the resolution with respect to the diffraction limit. The Strehl ratio shown in the figure is the one-dimensional Strehl ratio, which is the ratio of the area under the actual MTF to the area under the diffraction limited MTF [42,43].

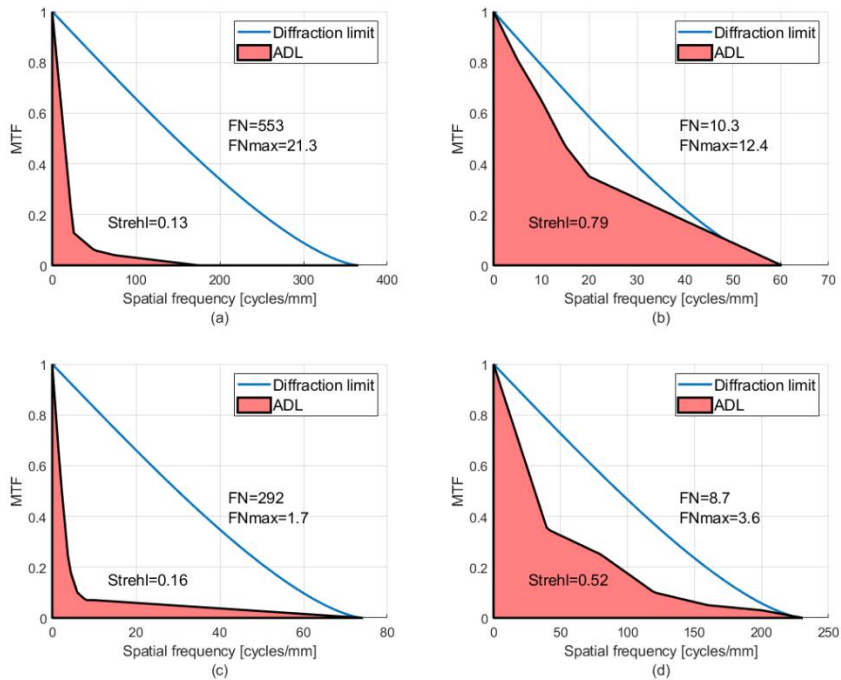

Fig. 1. MTFs of ADL designs compared to the diffraction limited MTF. For each graph the actual design FN is shown, compared to the FN upper limit presented in this paper. It can be seen that the larger the FN relative to the upper limit, the lower the resolution relative to the diffraction limit. (a-d) Designs 1,2,8, and 11 respectively.

## 6. Discussion

The upper limit on the Fresnel number for the different types of flat lenses may seem like a hard threshold, but in fact it provides much more than that. We have seen that a design can exceed this limit, but this will come at the expense of reduced Strehl ratio. Furthermore, the more the design exceeds the limit, the more severe the degradation in resolution, i.e. the lower the Strehl ratio. Therefore, to understand what type of design (CDL, AML or ADL) would be better for a certain application, we can look at the FN limit, and see which gives a higher limit. The higher the limit, the better the performance is expected to be, even if the limit is exceeded. Of course, other factors, such a manufacturing complexity, cost and other system requirements must be considered, but the FN limit is a good start. To better understand the implications of this, let us look at a test case example of a cell phone camera lens design.

We would like to design an achromatic flat lens for the visible spectral range, which we will define here as $\lambda_0$=550nm, $\Delta\lambda$=200nm. The FN upper limit for a CDL, based on Eq. 16, comes out to be 550/200=2.75. For a metalens the largest $\Delta n$ is about 1 (TiO$_2$ truncated waveguides in background of glass or polymer), and the maximum antenna height is on the order of the wavelength, i.e. $p \approx 1$. Substituting these values into Eq. 4 results in $FN_{max}$=5.5, i.e. double that of the CDL. Now let us look at an ADL. There we typically have $\Delta n \approx 0.65$ (when manufactured in photoresist) and $p \approx 6$. According to Eq. 10 we obtain $FN_{max}$=4.9. This is in the same ballpark as the AML. Of course, the AML/ADL can be somewhat improved by increasing $p$ or $\Delta n$, but this comes at the expense of more complex manufacturing.

Is this useful for common flat lens applications? If we calculate the Fresnel number of a typical cellphone camera lens ($f$=4mm, NA=0.2, $\lambda_0$=550nm) based on Eq. 3, we obtain FN=291, which is almost two orders of magnitude larger than the Fresnel numbers mentioned above. Thus, we can expect low resolution for such a design, like what is seen in Fig 1(a,c). Therefore, while AMLs and ADLs can provide an improvement over a CDL, they are still not the ultimate solution for most applications.

An important advantage of ADLs over AMLs is that they can have very large bandwidth. For example, if our CWL is 550nm, using the same $\Delta n$ and $p$ as before, the maximum wavelength for an ADL based on Eq. 12 is seven times the CWL, so we can have an ADL that covers a very large spectral range of about 400-3850nm (VIS-MWIR). This type of broad bandwidth ADL is demonstrated in recent work [27,28]. In essence, this is originated from the ability of an ADL to operate at moderately high $m$ values, whereas the implementation of such $m$ values in AMLs is extremely challenging due to the deep subwavelength features that are needed.

Based on Eq. 4 and 10, for AML and ADL respectively, we can conclude that AMLs are better suited (i.e. give a higher $FN_{max}$) for narrow bandwidth applications (where they can provide high FN with good resolution), while ADLs are better suited for large bandwidth applications (where they are limited to low FN, or alternatively to poor resolution at high FN, but nonetheless will perform better than an equivalent AML). By comparing Eq. 4 and 10 (assuming $m$>>1) we find that an ADL is better for spectral ranges satisfying $\Delta\lambda/\lambda_0 > 2$ (of course this is not an accurate cutoff, but gives an estimate of the situation).

Interestingly, in a recent paper that compares the diffraction efficiency of a chromatic truncated-waveguide type metalens to that of a chromatic diffractive lens, it was found that a metalens has an advantage in angular coverage, including both high-NA and large FOV, while a diffractive lens has an advantage in wavelength range coverage [44]. Based on this it may be possible to generalize the conclusion of the previous paragraph from achromatic metalenses and diffractive lenses to metalenses and diffractive lenses in general.

A well-known application for diffractive lenses is a refractive-diffractive doublet, for correction of chromatic aberration of refractive lenses. For this application, the upper limit on Fresnel number derived here is irrelevant, since the chromatic aberration is being used to advantage. In this context, an application for a dispersion engineered metalens has been demonstrated [45], that allows correction of the secondary chromatic aberration. For this type of application, it seems there is an advantage to metalenses, since the truncated waveguide dispersion can be tailored to need.

In our analysis, we focused on the case of continuous spectrum, which is typical for imaging systems. For discrete wavelength applications, the wavelength limits found in this paper apply, but not the FN limit. This is the case also for the AML upper limit presented in [26]. In the case of an AML we no longer need such high dispersion. It is sufficient to have a large enough library of dispersive truncated waveguide designs so the desired phase functions can be implemented at a few specific wavelengths [46]. For an ADL, if the discrete wavelengths are harmonics, we no longer have chromatic blur. For both cases, it seems that the fewer and more widely spaced the wavelengths, the easier the design (feature height $h$ can be smaller), allowing to achieve higher FN alongside with good resolution. As the number of wavelengths and their density increase, we approach the limits presented in this paper. Calculation of relevant limits for the discrete wavelength case, and comparison of AML to ADL in this case, are left for future research.

## 7. Conclusion

The field of flat lenses in general, and achromatic flat lenses in particular, is being actively researched. A lot of excellent work has been done on methods of correcting the chromatic aberration of flat lenses. While the limitations of the current AML technology have been discussed, it was unclear what the limitations of ADL technology are. The purpose of this paper is to bridge this gap. We have shown that current ADL technology has a strong limitation on the maximum Fresnel number that can be achieved while preserving a nearly diffraction limited resolution. This imposes a limit on the practical applicability of the technology for applications involving broadband illumination. It is our hope that this paper will help clarify some of the underlying physics, promote good engineering practice in the design of flat lenses, and provide a basis for innovative design ideas and applications.

### Funding

Israel Ministry of Science, Technology and Space.

### Disclosures

The authors declare no conflicts of interest.